\documentstyle[12pt]{article}
\newcommand{\be}{\begin{equation}}
\newcommand{\ee}{\end{equation}}
\newcommand{\ba}{\begin{eqnarray}}
\newcommand{\ea}{\end{eqnarray}}

\begin{document}
\begin{center}
 {\bf\Large{
  Hamilton-Jacobi formalism of the massive Yang-Mills theory revisited }}
\end{center}
\begin{center}

{\bf Dumitru Baleanu}\footnote[1]{ On leave of absence from
Institute of Space Sciences, P.O BOX, MG-23, R 76900
Magurele-Bucharest, Romania, E-mail: dumitru@cankaya.edu.tr},

\end{center}
\begin{center}
Department of Mathematics and Computer Sciences, Faculty of Arts
and Sciences, Cankaya University-06530, Ankara , Turkey
\end{center}
\begin{abstract}
Using Hamilton-Jacobi formalism we investigated the massive\\
Yang-Mills theory on both extended and reduced  phase-space.\\
The integrability conditions were discussed and the actions were
calculated.
\end{abstract}

\section{Introduction}

 Since Dirac initiated \cite{dirac} the Hamiltonian formulation of
constrained systems, the second class systems were subjected to an
intense debate \cite{const}. Based on \\
Caratheodory's idea \cite{car} Hamilton-Jacobi formalism (HJ) for
constrained systems was initiated \cite{g5,g6} and developed in
\cite{gb1,gb2,gb3,pime}. Specific for this formalism is the fact
that we started with many "Hamiltonians" and imposing the
integrability conditions we obtain the action. It was proved that
(HJ) stabilization algorithm gives the same results as Dirac's
procedure although there are two different ways of obtaining the
constraints. For second-class constrained systems (HJ) formalism
has many subtle aspects which should be clarified. First of all
the system corresponding to the
 total differential equations are not integrable. To make it
integrable we must modify it.The first option is to solve the
total differential equations by adding the new set of equations
obtained in the process of variation. This procedure is working
properly if the form of the equations is not so complicated and it
is extremely difficult to apply it if the equations are non-linear
ones.
 For these reasons new ways of making the system integrable should
be developed without solving the total differential equations. One
way is to transform the system, by reducing or extending the
phase-space, in such a manner that the modified "Hamiltonians"
have a physical interpretation from (HJ) point of view
\cite{rovelli}. In this paper we will present two methods of
making an integrable system from (HJ) point of view. The first way
is based on the chain method and the second one is deeply related
to the powerful formalism Batalin-Fradkin-Vilkovisky (BFT)
\cite{fad} or its modified versions \cite{barcelos-neto},
\cite{park}. To illustrate the methods we investigated the gauge
non invariant massive Yang-Mills theory, which is one of the
important models admitting non-linear second-class constraints.
\\
\\
\\
  The plan of the paper is as follows:

 In Sec. 2  (HJ) formalism and the chain method were briefly
 presented. In Sec.3  the massive Yang-Mills theory is analyzed within (HJ) formalism.
  Our conclusions are given in Sec. 4.

\section{Hamilton-Jacobi formalism}

Let us assume that the Lagrangian L is singular and the Hessian
supermatrix has rank n-r. The Hamiltonians to start with are

\be\label{doi} H_{\alpha}^{'}=H_{\alpha}(t_{\beta},q_{a},p_{a})
 +p_{\alpha}, \ee where $\alpha,\beta=n-r +1,\cdots,n$,$a=1,\cdots
n-r$. The usual Hamiltonian $H_0$ is defined as

\be\label{unu} H_{0}=-L(t,q_{i},{\dot q_{\nu}},{\dot q_{a}=w_{a}})
+p_{a}w_{a} + {\dot
q_{\mu}}p_{\mu}\mid_{p_{\nu}=-H_{\nu}},\nu=0,n-r+1,\cdots,n. \ee
which is independent of $\dot q_{\mu}$. Here $\dot
q_{a}={dq_{a}\over d\tau}$, where $\tau$ is a parameter.
 The equations of motion are obtained as total differential equations
in many variables as follows

\ba\label{(pq)} &dq_{a}&=(-1)^{P_a +P_a P_\alpha} {\partial_{r}
H_{\alpha}^{'}\over\partial p_{a}}dt_{\alpha},
dp_{a}=-(-1)^{P_{a}P_{\alpha}}{\partial_{r}
H_{\alpha}^{'}\over\partial q_{a}}dt_{\alpha},\cr
&dp_{\mu}&=-(-1)^{P_{\mu}P_{\alpha}}{\partial_{r}
H_{\alpha}^{'}\over\partial t_{\mu}}dt_{\alpha}, \mu=1,\cdots, r ,
\ea

\be\label{(z)} dz=(-H_{\alpha} + (-1)^{P_a + P_a
P_\alpha}p_{a}{\partial_{r} H_{\alpha}^{'}\over\partial
p_{a}})dt_{\alpha}, \ee where $z=S(t_{\alpha},q_{a})$ and $P_{i}$
represents the parity  of $a_{i}$.

On the surface of constraints the system of differential
equations(\ref{(pq)}) is integrable if and only if
\be\label{condh} [H_{\alpha}^{'},H_{\beta}^{'}]=0. \ee

 As we can say from (\ref{condh}) if the system is second class in Dirac's
classification, then it is not integrable. For second class
constrained systems some of the "Hamiltonians" are not in the form
(\ref{doi}) and some of them are not in involution. In the
following we will solve these problems using two different
techniques working in reduced phase-space and extended phase-space
respectively.

\subsection{The chain method}

  Let us assume
that the local Lagrangian density L is singular and admits only
one second-class constraints \cite{mitra} $\phi_{1}({\vec
x}),\phi_{2}({\vec x}),\cdots,\phi_{2n}({\vec x}) $ and only
$\phi_{1}({\vec x})$ is primary. In this method we have infinite
number of constraint equations. If we calculate the bracket
between constraints we will construct the matrix
$M_{ij}=\{\phi_i,\phi_j\}$ as

\be\label{matrice2} M_{ij}({\vec x},{\vec
y})=\left(\begin{array}{ccccccc}
0 & 0 & 0 & \cdots & 0 & 0 &  -\alpha\\
0 & 0 & 0 & \cdots & 0 & \alpha & *    \\
0 & 0 & 0 &         & -\alpha& * & *  \\
\vdots&  &         &        &   & & \vdots\\
0 & 0 & \alpha     &  & 0 & *&*    \\
0 & -\alpha& * &   &        * & 0 & * \\
\alpha&*   & * &\cdots&*      &*& 0\\
\end{array}\right)\delta({\vec x}-{\vec y}),
\ee where $\{\phi_1,\phi_{2n}\}=-\alpha$.

 To simplify  the problem  and to explain the method we consider
the two-chain case. Let us assume that the set of the constraints
in involution is given
$\{\chi_1,\chi_2,\cdots,\chi_r;\psi_1,\psi_2,\cdots, \psi_{q}\}$.
To eliminate the other constraints we transform the Hamiltonian as
\cite{mitra} \be\label{extra} H^{''}=H_c+{1\over
2}\chi^{T}\alpha^{-1}\chi, \ee where

\be \chi=\left(\begin{array}{c} \chi_{r+1} \\
\psi_{q+1}\end{array}  \right). \ee

Let us  assume that for field theory only a primary constraint
$\Phi_1$ generates 2n-1 constraints denoted by $\Phi_{\alpha},
\alpha=2,\cdots 2n$. If the corresponding extended Hamiltonian has
the following form  \be\label{extra1} H^{'''}= H_c +{1\over 2}\int
d{\vec x}\alpha^{-1}({\vec x})\Phi_{n+1}^{2}(\vec x), \ee then
 half of the constraints are eliminated and the resulting
system is a first-class one (for more details see
Ref.\cite{mitra}).

  If, for a given second-class constrained system we will find a
subset of constraints in involution and in the form
(\ref{doi}),then due to (\ref{condh}) the system is integrable
from (HJ) point of view and we can calculate the action.

\section{Massive Yang-Mills theory}

 The Lagrangian density is given by \be\label{lcan} L=-{1\over
4}F_{\mu\nu}^{a}F^{a\mu\nu}+{1\over 2}m^2 A_{\mu}^a A^{a\mu},\ee
where \be
F_{\mu\nu}^a=\partial_{\mu}A_{\mu}^a-\partial_{\nu}A_{\mu}^a
+gf^{abc}A_{\mu}^b A_{\nu}^c \ee In addition we use the following
notations and conventions \be [T^a, T^b]=if^{abc}T^c,
tr(T^aT^b)={1\over 2}\delta^{ab}, (T^a)^{bc}= i f^{abc}. \ee The
canonical Hamiltonian density corresponding to (\ref{lcan}) is
\be\label{hcan} H_c={1\over 2}(\pi_i^{a})^2+\pi^{ai}{\partial_i
A_0^a}-gf^{abc}\pi^{ai}A_0^bA_i^c +{1\over
4}F_{ij}^{a}F^{aij}-{1\over 2}m^2A_i^aA^{ai}-{1\over
2}m^2A_0^aA^{a0}.\ee

 Let us apply the chain method to investigate the constraints of
the system. The primary constraint is \be \phi_1^a=\pi_0^a=0 \ee
and the total Hamiltonian is given by \be H_T = \int
d^3x(H_c+\lambda^a\pi_0^a), \ee where $\lambda^a$ are Lagrange
multipliers. The consistency condition of $\phi_1^a$ gives another
constraint \be \phi_2^a=(D_i\pi^i)^a +m^2A_0^a, \ee where
$(D_i\pi^i)^a={\partial_i\pi^{ai}}+gf^{abc}A_i^b\pi^{ci}$. If we
impose the consistency condition for $\phi_1^a$ we obtain an
equation for $\lambda^a$. The system is second-class and the
Poisson algebra of $\phi_0^a$ and $\phi_1^a$ is given by \be
\begin{array}{c}
 \{\phi_0^a(x), \phi_0^b(y)\}=0,\{\phi_0^a(x),
\phi_1^b(y)\}=-m^2\delta^{ab}\delta({\vec x}-{\vec
y}),\\\{\phi_1^a(x),
\phi_1^b(x)\}=gf^{abc}(D_i\pi^i)^c\delta({\vec x}-{\vec y}).
\end{array}
\ee
 In (HJ) formalism the "Hamiltonian" densities to start with are
\be\label{ppr} H_0^{'}=p_0+H_c, H_1^{'a}=\pi_0^a. \ee If we
consider the variations of $H_1^{'a}$ we obtain another
"Hamiltonian" \be\label{fata} H_2^a= (D_i\pi^i)^a +m^2A_0^a.\ee
The variation of (\ref{fata}) gives an equation for $\dot A_0^a$
which is nothing than the form of Lagrange multipliers $\lambda^a$
from the previous method. From (\ref{ppr}) and \ref{fata} we
deduce that the "Hamiltonians" are not in the form required by
(\ref{condh}). At this stage we will transform the "Hamiltonians"
such that they will be in involution on the reduced phase-space.
We choose $H_1^{'a}=\pi_0^a$ and we transform $H_0^{'}=p_0+H_c$ as
follows \be\label{hyt} {\tilde H_0^{'}}= p_0+H_c -{1\over
2}{1\over m^2}\int\left(d{\vec
x}\left[\sum_{a=1}^{a=n}(D_i\pi^i)^a +m^2A_0^a\right]^2\right).\ee
In this manner we obtain an integrable system corresponding to
${\tilde H_0^{'}}$ and $H_1^{'a}$. The corresponding action of
these two "Hamiltonians" is given by \be
\begin{array}{c}
S=\int{\{{-1\over 4}F_{ij}^{a}F^{aij}+{1\over
2}\pi^{ai}\pi_i^a+{1\over 2}m^2A_0^aA^{a0}-{1\over
2}m^2A_i^aA^{ai}}\\
-{1\over 2m^2}\sum_{a=1}^{a=n}[(D_i\pi^i)^a
+m^2A_0^a][(D_i\pi^i)^a -m^2A_0^a]\} d{\vec x}.
\end{array}
\ee

\subsection{Modified BFT conversion}

 Another method to make the "Hamiltonians" in involution and to keep their
physical significance from (HJ) point of view is based on the use
of (BFT)
 formalism \cite{fad}.
 Let us suppose that for a system with N degrees of freedom $T_a$
second-class constraints exists, where $\quad a=1,\cdots, M < 2N$
and \be \{T_a,T_b\}=\Delta_{ab} \ee with $det(\Delta_{ab})\neq 0$.
The aim of the formalism  is to transform the above constraints
into first-class ones by adding auxiliary variables $\eta^{a}$
\cite{fad}. In addition the auxiliary variables fulfills a
symplectic algebra \be \{\eta^a,\eta^b\}=\omega^{ab}, \ee where
$\omega^{ab}$ represents a constant quantity subjected to the
restriction $det(\omega^{ab})\neq 0$. For (HJ) formalism the form
of the first-class constraints ${\hat T_a}={\hat T_a(q,p,\eta)}$
is crucial.In \cite{fad} the form of the constraints in involution
was required to be \be {\hat T_a}=\sum_{n=0}^\infty T_a^{(n)},\ee
where $T_a^{(n)}$ represents the $n^{th}$ order in q and $\eta$.
To keep the physical significance of the "Hamiltonians" we require
them to be in the form \be\label{ght} {\hat H_\alpha^{'}}={\hat
p_\alpha}+{\hat H_\alpha} \ee
 on the extended phase-space. In (\ref{ght}) the involved quantities are functions on the extended phase-space.
  The key point is to calculate the first correction
 $T_a^{(1)}=X_{ab}(q,p)\eta^{b}$ in such a way to obtain  ${\hat T_a}$ in the form given by (\ref{ght}).
 Taking into account that for the massive Yang-Mills theory we have
\be \omega_{A,B}^{ab}( x, y)=\left(\begin{array}{cc}
0 & 1\\
-1 & 0
\end{array}\right)\delta^{ab}\delta( x- y).
\ee
 after calculations we found that
 \be X_{AB}^{ab}( x, y)=\left(\begin{array}{cc}
m^2\delta^{ab} & 0\\
-{1\over 2}g f^{abd}(D_i\pi^i)^d & \delta^{ab}
\end{array}\right).
\ee
 Following (BFT) formalism we found the constraints in
involution as

 \be
  {\hat \phi_1}^a=\pi_0^a+m^2\eta^{1a},\\
{\hat \phi_2}^a=\eta^{2a}+(D_i\pi^i)^a +m^2A_0^a
-\sum_{n=1}^{\infty}[(g\bar\eta)^n]^{ab}(D_i\pi^i)^b,
 \ee
 where
 \be \bar\eta^{ac}=f^{abc}\eta^{1 b}. \ee
 Using the technique described in \cite{barcelos-neto, park}
 we will construct the involutive forms of the initial fields.
 After a tedious calculations we obtained
 \be
 \begin{array}{c}
   {\hat\pi_i^{a}}=\pi_i^{a}+\sum_{n=1}^{\infty}(-1)^n{g^n\over n!}(\bar\eta^{n})^{ac}\pi_i^c, \\
   {\hat A_i^a}=A_i^a +{1\over (n+1)!}\sum_{n=1}^{\infty}
({g\bar\eta}^{n})^{ac}D_i\eta^{2c}\\
   {\hat A_0^a}=A_0^a +{1\over
m^2}\sum_{n=1}^{\infty}[\eta^{2a}\delta_{1n}+b_n(({g\bar\eta}^{n})^{ac})(D_i\pi^{i})^c],\\
 \end{array}
 \ee
 where \be b_n={1\over n}[-b_{n-1}+{1\over (n+1)!}], n> 1,
b_1={1\over 2}. \ee We mention that the our results are different
from those obtained in \cite{barcelos-neto, park}.
 The involutive Hamiltonian ${\hat H_c}({\hat A_0},{\hat
A_i},{\hat\pi_i^a})$ is obtained from
 ${H_c}({A_0},{A_i},{\pi_i^a})$
 by making the following substitutions $A_0\rightarrow{\hat A_0},
A_i\rightarrow{\hat A_i},\pi_i^a\rightarrow{\hat \pi_i^a}$.
 "Hamiltonians"
$H_1^{'a}={\hat \phi_1}^a,H_2^{'a}={\hat \phi _2}^a$ and
$H_0^{''}=p_0+{\hat H_c}$ are in involution and all of them have
physical significance from (HJ) point of view \cite{rovelli}. The
corresponding action of $H_1^{'a},H_2^{'a}$ and $H_0^{''}$ is
given by \be
\begin{array}{c}
 dS=-m^2\eta^{1a}d\pi^{0a}-(m^2A^{a}_0)d\eta^{1a}+\{{1\over
2}(\pi_i^{a}+\sum_{n=1}^{\infty}(-1)^n{g^n\over
n!}(\bar\eta^{n})^{ac}\pi_i^c)^2\\
(\pi_i^{a}+\sum_{n=1}^{\infty}(-1)^n{g^n\over
n!}(\bar\eta^{n})^{ac}\pi_i^c)\left({\partial_i}[b_n({g\bar\eta}^{n})^{ac}]gf^{cdb}{A_j^d}\pi^{jd}-
\partial_j^2[b_n({g\bar\eta}^{n})^{ac}]\pi^{jc}\right)\\
-{1\over 4}F_{ij}^{a}F^{aij}+{1\over 2}m^2A_i^aA^{ai}
+gf^{abc}\pi^{ai}A_0^bA_i^c+\\
(A_0^a + {1\over
m^2}\sum_{n=1}^{\infty}[\eta^{2a}\delta_{1n}+b_n(({g\bar\eta}^{n})^{ac})(D_i\pi^{i})^c])
[{1\over 2}A_0^a- {1\over
2m^2}(\sum_{n=1}^{\infty}(-\eta^{2a}\delta_{1n}+b_n(({g\bar\eta}^{n})^{ac}))]\}dt.
\end{array}
\ee

\section{Conclusions}

 Second-class constrained systems are problematic for (HJ)
formalism because the corresponding  system of total differential
equations are not integrable. In this paper we presented another
way to make the "Hamiltonians" in involution, and implicitly to
obtain an integrable system, without solving the equations of
motion. The basic idea was to keep the physical significance of
the modified "Hamiltonians".
 In the first part
we applied the chain method to make the system corresponding to
the massive Yang-Mills theory integrable on the reduced
phase-space. Due to the form of the modified constraints all
"Hamiltonians" have physical significance from (HJ) point of view.
In the second part we used a method based on modified (BFT)
formalism and we obtained an integrable system on the extended
phase-space. In (BFT) formalism the process of making the
constraints in involution is not unique. Using this important
property of this approach we followed that way which gives the
physical significance for all "Hamiltonians" on the extended
phase-space. The obtained results are different from those
presented in \cite{barcelos-neto, park}. In both cases the action
corresponding to the involutive "Hamiltonians" was calculated .

 \section {Acknowledgments}
 The author thanks M. Henneaux for encouragements.This work is partially supported by
the Scientific and Technical Research Council of Turkey.

\end{document}